\documentclass{article}

\usepackage{xcolor}
\selectcolormodel{gray}

\usepackage{amssymb}
\usepackage{pgfplots}
\usepackage{xy}
\usepackage{caption}
\usepackage{floatrow}
\newfloatcommand{capbtabbox}{table}[][\FBwidth]
\usepackage{blindtext}

\usepackage[showframe=false, margin = 1.5 in]{geometry}
\usepackage[shortlabels]{enumitem}
\usepackage{cite}
\usepackage{tikz, xypic}
\tikzstyle{vertex} = [fill,shape=circle,node distance=80pt]
\tikzstyle{edge} = [fill,opacity=.5,fill opacity=.5,line cap=round, line join=round, line width=50pt]
\tikzstyle{elabel} =  [fill,shape=circle,node distance=30pt]

\pgfdeclarelayer{background}
\pgfsetlayers{background,main}
\usepackage{comment}
\usepackage{color}

\usepackage{hyperref}
\hypersetup{colorlinks,linkcolor={red},citecolor={olive},urlcolor={red}}
\usepackage{mathtools}
\mathtoolsset{showonlyrefs}

\usepackage{multicol}

\usepackage{theoremref}

\newcommand{\f}{\frac}
\renewcommand{\L}{\mathcal L}

\newcommand{\M}{\mathcal M}
\renewcommand{\H}{\mathcal H}

\title{Safe reopening of university campuses is possible with COVID-19 vaccination}
\date{}

\author{Matthew Junge \thanks{Department of Mathematics, Baruch College. This work is partially supported by NSF grants DMS-1855516 and RAPID-2028892.} 
\and 
{Sheng Li} \thanks{Department of Epidemiology, School of Public Health, City University of New York. Partially supported by the CUNY SPH Dean's COVID-19 Grant and IRG Multidisciplinary Research Grant.} 
\and 
Samitha Samaranayake \thanks{School of Civil and Environmental Engineering, Cornell University. This work is partially supported by NSF grant CMMI-2033580 and US-DOT grant 69A3551747119.}
\and
Matthew Zalesak \thanks{School of Operations Research and Information Engineering, Cornell University. This work is partially supported by NSF grant CMMI-2033580 and US-DOT grant 69A3551747119.}
}

\begin{document}

\maketitle

\begin{abstract}
We construct an agent-based SEIR model to simulate COVID-19 spread at a 16000-student mostly non-residential urban university during the Fall 2021 Semester. We find that mRNA vaccine coverage above 80\% makes it possible to safely reopen to in-person instruction. If vaccine coverage is 100\%, then our model indicates that facemask use is not necessary. Our simulations with vaccine coverage below 70\% exhibit a right-skew for total infections over the semester, which suggests that high levels of infection are not exceedingly rare with campus social connections the main transmission route. Less effective vaccines or incidence of new variants may require additional intervention such as screening testing to reopen safely. 
\end{abstract}

\section{Introduction}

 During March 2020 of the COVID-19 pandemic most universities in the United States halted on-campus operations and went to remote instruction. About one third reopened to full, or partial in-person instruction in  Fall 2020, with more schools following suit in the Spring \cite{yamey2020covid, spring_reopening}.  Many reopenings have been accompanied by infection spikes which required temporary pivots to remote instruction \cite{wilson2020multiple, college_superspreader}.
 As of April 30 2021, the New York Times reports over 660,000 confirmed cases of COVID-19 on college campuses. These are directly linked to over 100 deaths, mostly involving employees \cite{nyt}.
 
 Human daily behavioral contracts are primary among similar age groups and secondarily with other age group family members. A big challenge introduced by reopening schools is the restoration of same age social contacts. A previous study has shown school closure and social distancing dramatically reduce daily physical contacts, particularly the dominant same age group contacts. This efficiently prevents the COVID-19 transmission \cite{zhang2020changes}.  
 
  Universities that reopened to in-person instruction implemented protocols to help control infection spread such as: periodic testing, mandatory facemask use, social distancing, building closures, limited extracurricular activities, and hybridized in-person/remote classroom instruction. As these levels of intervention lacked much precedent, various models were developed to help guide policy and predict outcomes \cite{gressman2020, bahl2020modeling}. Policy decisions ultimately struck a balance between forecasts, campus safety and comfort, and university resources \cite{reasons}.

 With the introduction of apparently effective vaccines \cite{jj, moderna, pfizer, chagla2021bnt162b2} and increased natural immunity from earlier exposure \cite{bajema2020estimated}, more universities are planning to conduct Fall 2021 primarily in-person \cite{NYT2}. Like in Fall 2020, it is still uncertain how much intervention is needed to control COVID-19 infections. We develop an agent-based SEIR model to forecast total COVID-19 infections over the course of a semester at a primarily non-residential urban university campus with 16000 students and 800 faculty. 
Baruch College, part of the over 275,000 student City University of New York (CUNY), is used to inform our framework. 

 Urban universities, such as those in the CUNY system, are usually located in densely populated areas and serve many students from minority groups. Preliminary surveys indicate that vaccine hesitancy from minority groups that present higher COVID-19 infection incidence and higher than average vaccine hesitancy will limit vaccine coverage to somewhere between 60-80\% of the the United States population with African Americans among the most hesitant \cite{dror2020vaccine, khubchandani2021covid, callaghan2020correlates, karpman2021confronting, razai2021covid}. 
 As many students live with their families, reopening such universities is accompanied by elevated risk to and from their households and communities.

 Depending on the vaccine administered, current clinical trials suggest efficacy ranging from 65--95\% \cite{jj,moderna, pfizer}. As these statistics are derived by comparing the symptomatic case incidence in the vaccination group to that in the placebo group, it is likely that asymptomatic cases are missed in these statistics. Preliminary data suggests that vaccination reduces asymptomatic cases as well \cite{tande2021impact, cdc, hall2021effectiveness, tang2021asymptomatic}.  The data in \cite{hall2021effectiveness} was obtained from biweekly testing in healthcare workers and the author's found the BNT162b2 vaccine 86\% effective at preventing symptomatic and asymptomatic spread. Another study \cite{tang2021asymptomatic} analyzes biweekly screening tests in a group of employees at St Jude's Childrens Hospital. A 72\% reduction in asymptomatic cases was observed. 
 The authors point out that short followup time, small cohort size, and that individuals choosing to not vaccinate might be higher risk could limit the accuracy of their findings. 
 It remains unclear to what extent vaccinated individuals can spread COVID-19, and how well the vaccines protect against variant strains of COVID-19.
 
 
\section{Methods}

  We model different scenarios with two primary variables: \emph{vaccine effectiveness} and \emph{vaccine coverage} of the campus population. We utilize the agent-based campus Susceptible-Exposed-Infected-Removed model from \cite{zalesak2020seir}. 
  Similar to \cite{gressman2020, bahl2020modeling}, students and faculty are assigned individualized schedules that they follow throughout a simulated semester. Schedules are organized into common meetings---classroom, broad environment, clubs, residential, socializing---during which COVID-19 is equally likely to be passed from infected agents present to each susceptible agent also present. The infection rate in our model is set to obtain an \emph{average reproduction number} $R_0=3$, which represents the average number of infections caused by an exposed agent in an entirely susceptible population.
  Estimates for COVID-19 spread in large communities (such as cities and countries) put $R_0$ in the range $[2.0,3.0]$ \cite{rahman2020basic, abbott2020transmissibility, ueki2020effectiveness, zhou2020preliminary, zhao2020preliminary}, but there are some higher estimates \cite{sanche2020early}. The statistic varies by community contact structure. It has been observed that $R_0$ is larger in reopened universities \cite{college_superspreader}. For example, \cite{gressman2020} sets $R_0=3.8$ in their campus COVID-19 model for a mostly residential urban university.  See  \cite{gressman2020} and \cite{bahl2020modeling} for more discussion about elevated $R_0$ levels in a university setting. We remark that our model implicitly assumes agents are wearing facemasks while on campus. The assumption is implicit because we base our choice of $R_0$ on what occurred in universities for the 2020-2021 school year in which, to our knowledge, all universities required facemasks in public spaces.

  Vaccination and antibody status impact agents' susceptibility and infectiousness. Each agent is assigned the \emph{vaccinated attribute} independently with probability $V$. Such agents have \emph{inward protection} factor $r_i$ and \emph{outward protection} factor $r_o$. Vaccinated agents contribute a factor of $1-r_o$ of exposure time to susceptible agents in each meeting they are present at. When computing the probability vaccinated agents are infected at the end of a day, the probability is multiplied by $1-r_i$. All COVID-19 infections in vaccinated agents are classified as asymptomatic. Our 80\%  vaccine effectiveness scenarios sets $r_i = 0.8 = r_o$, while 50\% effectiveness has $r_i=0.5=r_o$.  The model is initiated with 10 randomly selected students in the exposed state. The main statistic we consider is the total number of agents ever in the exposed state. We refer to this as \emph{total infections}.

\section{Results} \label{sec:overview}


          
            

 \begin{figure}
          \begin{floatrow}
             \ffigbox{\includegraphics[width = .49 \textwidth]{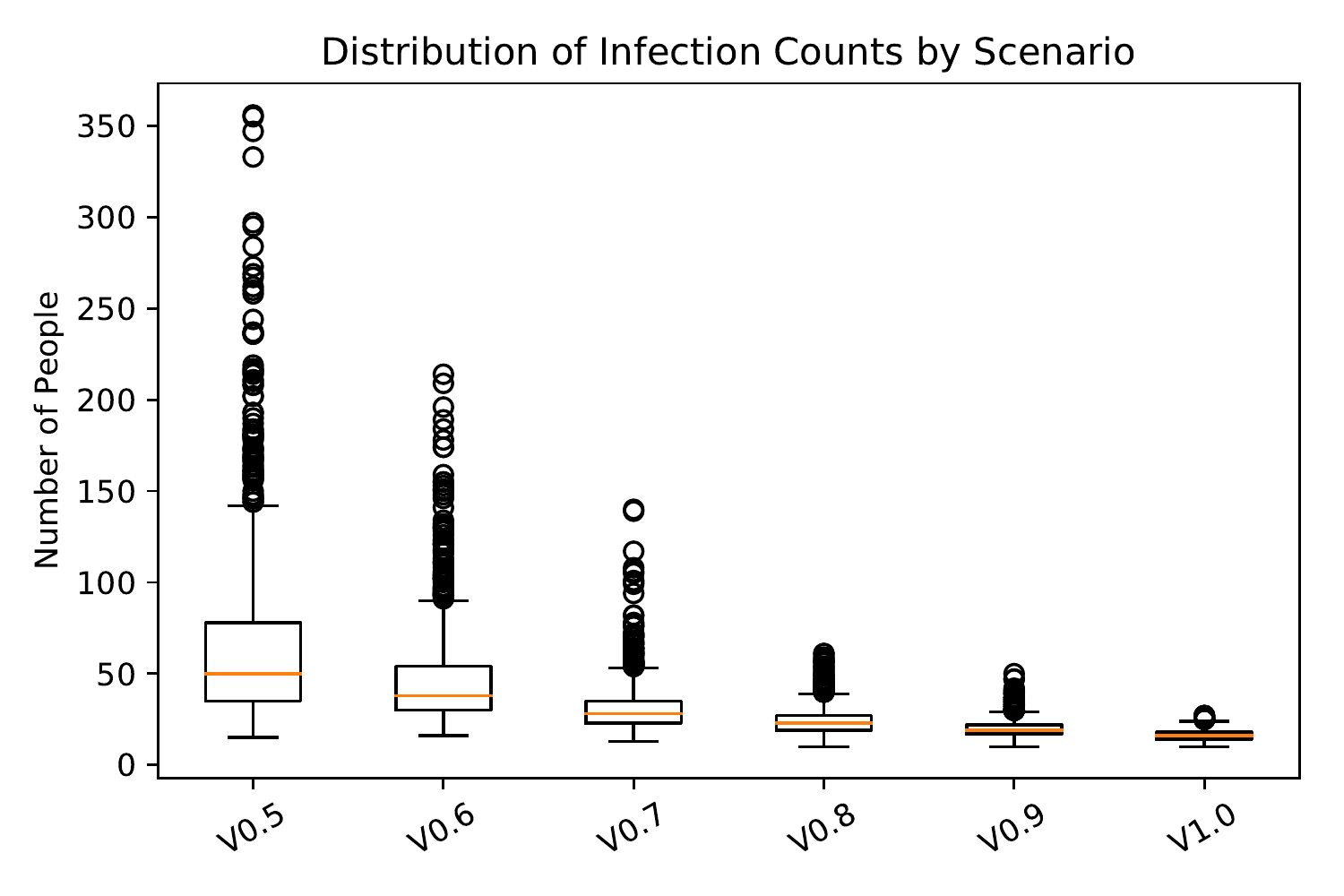}}{\caption{Total infections ($y$-axis) in 1000 simulated 15-week semesters per scenario with 80\% vaccine effectiveness. Each scenario has a proportion $V \in \{0.5,0.6, \hdots, 1.0\}$ of the population vaccinated ($x$-axis).}\label{fig:vax81}}
             \ffigbox{\includegraphics[width = .49 \textwidth]{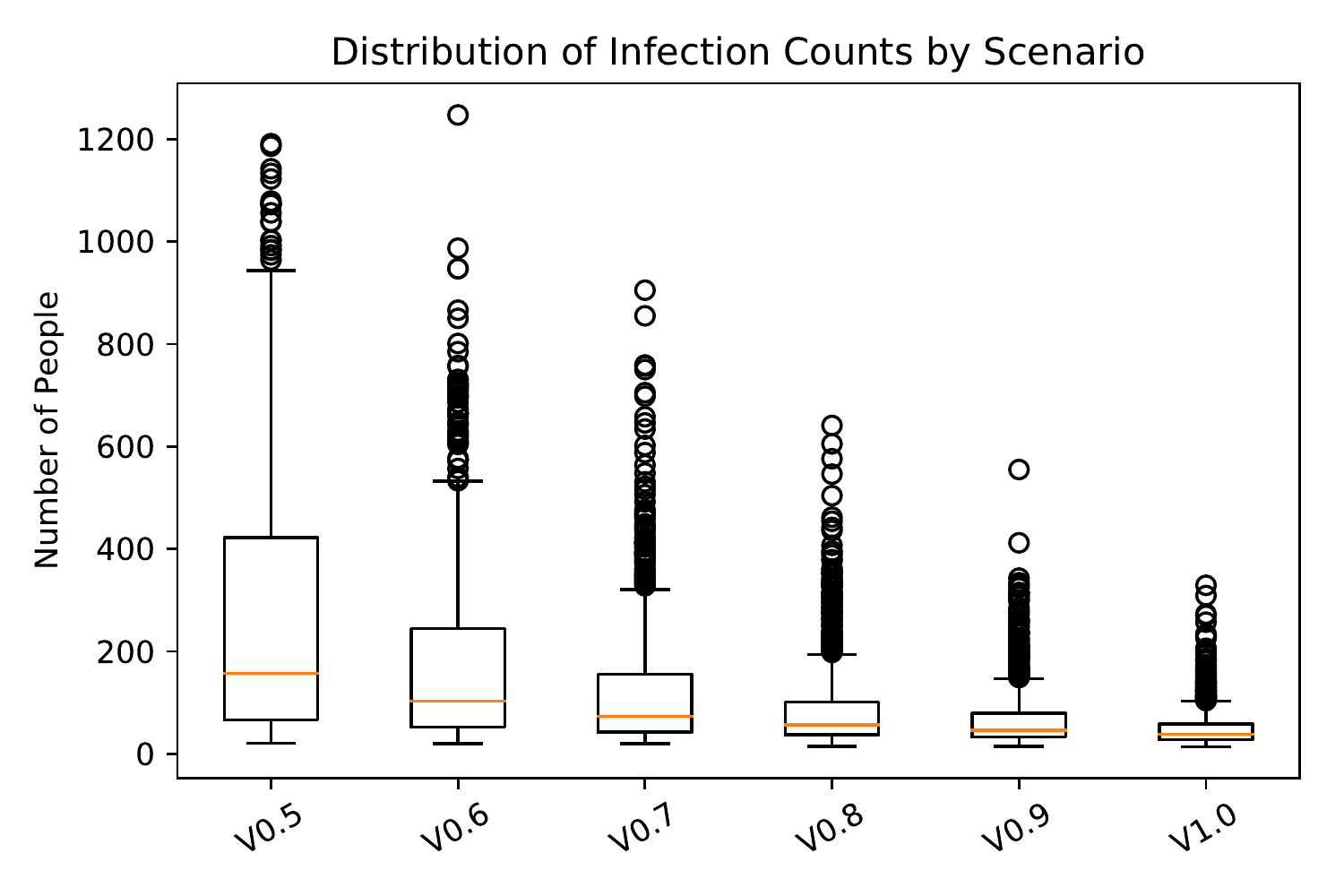}}{\caption{The same scenarios as in Figure~\ref{fig:vax81}, but with 50\% effectiveness for the vaccine.}\label{fig:vax51}}
          \end{floatrow}
\end{figure}

Figure~\ref{fig:vax81} displays box-plots with medians for the total number of infections in 1000 simulated semesters given different levels of vaccine coverage $V \in \{0.5, 0.6,\hdots, 1.0\}$. In these simulations, vaccinated individuals are assumed to have an 80\% reduction in the probability of becoming infected and a 80\% reduction in the probability of infecting others. This is consistent with current data for the mRNA COVID-19 vaccines \cite{tande2021impact, hall2021effectiveness}. We see that the infection is well controlled with vaccination levels above 80\%. A median 30 total infections occur at $V=0.8$. 
Although the median is below 50 for all $V \geq 0.5$, lower levels of vaccine coverage have right-tail events with relatively high numbers of total infections. We see simulations with 350 total infections when $V=0.5$ and over 150 total infections when $V=0.7$.

The plots in Figure~\ref{fig:vax51} have the same response and independent variables as in Figure~\ref{fig:vax81}, but display total infections with 50\% vaccine effectiveness.
While increasing vaccine coverage reduces total infections, a striking feature of the data is that all scenarios exhibit right-skew. We see multiple outliers; some simulations in Figure~\ref{fig:vax51} have over 1200 total infections. This indicates nearly 6\% of the campus population becoming infected, which is much higher than current national and local COVID-19 incidence. Even with 100\% of the campus population vaccinated, there are simulations with nearly 200 total infections (over 1\% of the campus population). 

\begin{figure}
          \begin{floatrow}
             \ffigbox{\includegraphics[width = .49 \textwidth]{scenario5_0.5-counts.pdf}}{\caption{Total infections by source with 50\% vaccine effectiveness and 25\% of the population screened for COVID-19 weekly.}\label{fig:testing51}}
             \ffigbox{\includegraphics[width = .49 \textwidth]{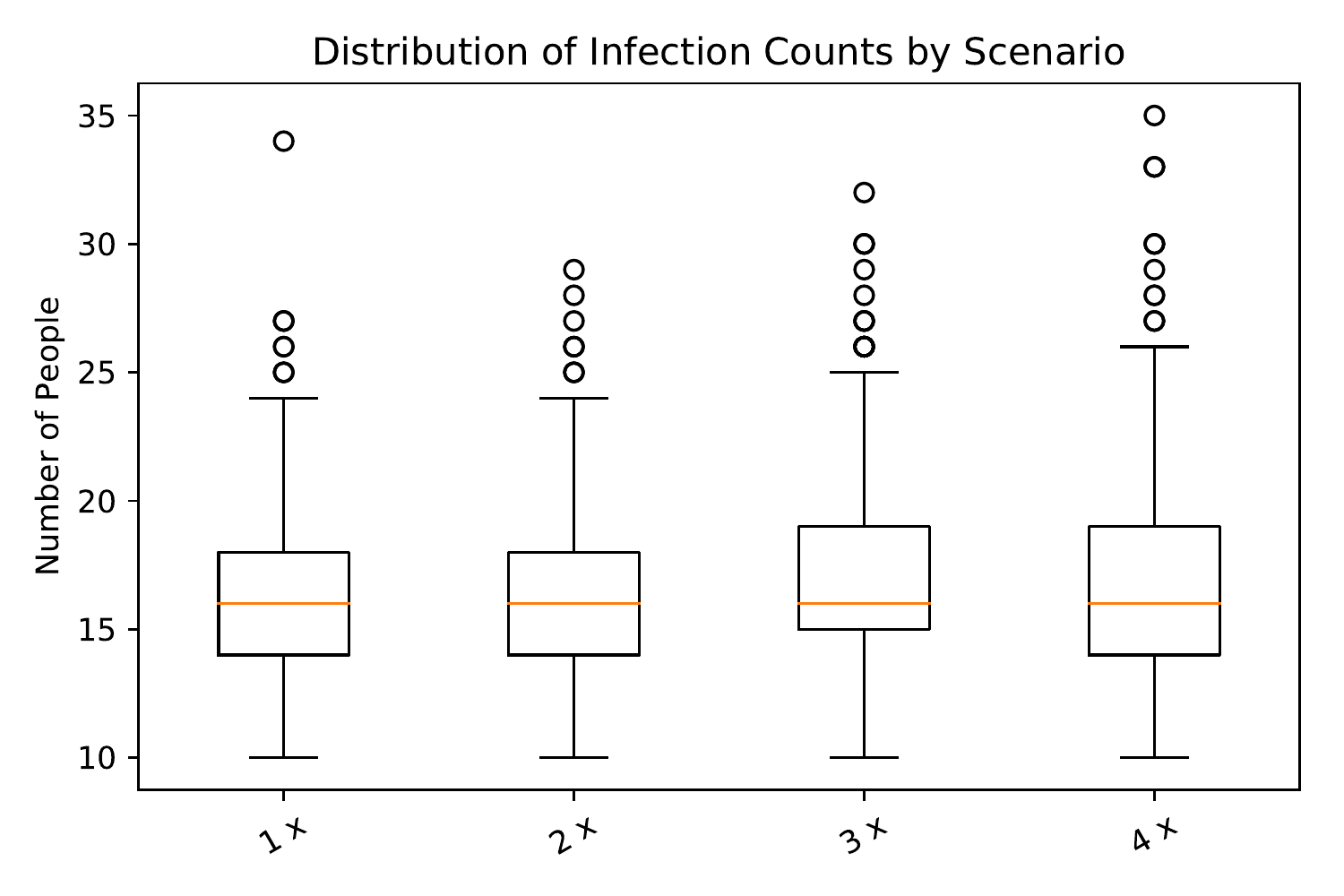}}{\caption{Total infections with 80\% vaccine effectiveness and 100\% vaccine coverage. To model no facemask use we increase the duration (intensity) of time spent in non-social settings by factors of 1, 2, 3, and 4.}\label{fig:fm}}
          \end{floatrow}
\end{figure}

Weekly screening of 25\% of the campus population reduces total infections. Figure~\ref{fig:testing51} 
displays total infections in the scenarios from Figure~\ref{fig:vax51}, but with 25\% of the campus population screened weekly for COVID-19. Individuals who test positive quarantine until recovering. A right-skew is still present, but it is less extreme than with no testing. Moreover, the medians for total infections in Figures~\ref{fig:testing51} are lower than in the analogous scenarios without testing  (Figure~\ref{fig:vax51}). 

Figure~\ref{fig:fm} shows simulations concerning the necessity of facemasks. We assume that the vaccine is 80\% effective and coverage is 100\%. Recall that our usual model assumes that facemasks are worn at all times in non-social settings.  It has been estimated that facemasks, when worn by both infected and susceptible agents, reduce the susceptible agents exposure to the virus by at least 50\% \cite{facemask, mask2}. To model no facemask use, we increase the duration of time spent in classrooms, the broad environment, clubs, and residence halls by factors of 2,3, and 4. Tripling the duration, for example, triples the probability agents become infected in each interaction in these spaces. We observe that these increases do not cause for many more total infections. The median total infections remains in the interval [15,20] and right tail events do not amount to many additional infections. In fact, over 1000 simulations we observed similar maximum total infections (34 versus 35) in the model with no multiplying factor and that with the risk multiplied by 4.    

\section{Discussion} \label{sec:rec}

On May 10, 2021, shortly before completing the analysis in this article, New York State Governor Andrew Cuomo announced that the City University of New York and State University of New York campuses will require proof of vaccination for all students attending in-person classes \cite{cuomo}. Our study suggests that, so long as vaccine efficacy is at or above 80\% for preventing both symptomatic and asymptomatic infections, this policy allows for a safe university reopening with minimal extra precautions needed. Our model typically assumes that individuals use facemasks while participating in campus activities. However, additional simulations suggest that facemasks are not necessary with 80\% vaccine effectiveness and 100\% vaccine coverage.

If the vaccines are less effective, then our results suggests that caution is needed. Even with high levels of vaccine coverage, vaccinated members of the campus population may still become infected.  Moreover, the right skew we observe for total infections implies that the risk of many agents in the model becoming infected is non-negligible. 
Most infections in our model are asymptomatic cases. The reason for this is twofold. First, asymptomatic cases are more common in young people \cite{galanti2019rates, davies2020age}. Secondly,  COVID-19 vaccines have been demonstrated to reduce symptomatic cases more effectively \cite{pfizer, moderna, jj, xiao2021early}.
 CUNY campuses have a large proportion of minority groups. The 2019 CUNY Student Data Book reports a 25\% Black and 30\% Hispanic undergraduate population across all 25 CUNY campuses. Minority groups exhibit more vaccine hesitancy \cite{karpman2021confronting} and neighborhoods with more minority residents have higher COVID-19 incidence \cite{nyc_tracker}.
 As many students at a non-residential urban university live at home, high levels of asymptomatic cases pose a silent risk to their households and communities. 

Ensuring that a large proportion of the campus population is vaccinated, and administering screening testing are effective ways to control total infections. Most of the infections in our model are from socializing. Accordingly, students should be encouraged to practice safe social contact during the semester such as distancing and wearing facemasks in the presence of unvaccinated students. This is similar to what was suggested in \cite{bahl2020modeling}.We further comment that infections will lower at least proportional to the level of dedensification employed by the university. For example, if half as many students are regularly on campus we expect that total infections will reduce by at least half.

A limitation with our model is the difficulty with setting parameters. Another is in designing the contact structure and relative risk of different meeting types. Socializing plays a dominant role for infection spread in our model, but it is difficult to create a realistic contact structure. A novel aspect of our approach compared to other agent-based COVID-19 university models \cite{bahl2020modeling, gressman2020} is that we use a Markov chain to create social groups matching students with similar characteristics (year and area of study). Our sensitivity analysis also includes scenarios with less socializing. On a different note, if screening testing is present on campus, then there is the opportunity for college administrators to respond in real-time to rising case counts. For example, moving to all remote instruction when a certain threshold is reached. For that reason, testing may be even more effective than our simulations suggest.

\section{Conclusion}

We constructed an agent-based SEIR model made to resemble a mostly non-residential urban university campus. We then ran different scenarios for vaccine effectiveness and coverage by the campus population. If the vaccines are 80\% effective at preventing both symptomatic and asymptomatic COVID-19 infections, then our study suggests that no extra precautions are needed for a safe reopening. Lower vaccine effectiveness and coverage may lead to undesirable levels of COVID-19 infections. A right-skew for total infections suggests that rare but extreme events could have particularly bad outcomes. Screening testing helps control total cases.  \\

 \section{Disclaimer} 

The contents of this report reflect the views of the authors, who are responsible for the facts and the accuracy of the information presented herein. This document is disseminated in the interest of information exchange. The report is funded, partially or entirely, by a grant from the U.S. Department of Transportation’s University Transportation Centers Program. However, the U.S. Government assumes no liability for the contents or use thereof.

\bibliographystyle{unsrt}
\bibliography{CUNYcovid}

\newpage 
\section{Appendix}

\subsection{Model specifics}

  We utilize the agent-based campus Susceptible-Exposed-Infected-Removed model from \cite{zalesak2020seir}. 
  Similar to \cite{gressman2020, bahl2020modeling}, students and faculty are assigned individualized schedules that they follow throughout a simulated semester. Schedules are organized into common meetings---classroom, broad environment, clubs, residential, socializing---during which COVID-19 is equally likely to be passed from infected agents present to each susceptible agent also present.

  All agents in the model start in either the susceptible state or with antibody protection. At the onset of the model, we independently assign each agent the \emph{antibody attribute} with probability $.20$. A proportion of the agents with this attribute have \emph{antibody protection} which prevents infection. Those without antibody protection act as normal susceptible agents.  If a susceptible agent becomes exposed to COVID-19 then, after a random \emph{incubation period}, the agent progresses to the asymptomatic or symptomatic infected state with equal probability \cite{galanti2019rates, davies2020age}. Such agents occupy this state for a random \emph{infectious period} and subsequently transition to the recovered state. Symptomatic individuals decide after a random \emph{observation period} to self-quarantine, either voluntarily or from seeking independent testing, until recovered. Recovered agents cannot become infected again.  Except for the quarantine period, periods are modelled with independent geometric random variables. We write Geo$(1/p)$ to denote the geometric distribution $P(X=k) = (1-p)^{k-1}p$ for integers $k \geq 1$ and $0<p<1$ which has mean $1/p$.

  \begin{figure}[h]
      \begin{equation} \xymatrix{ & & & & \fbox{$I_a$}  \ar_5[rd] \\
\fbox{$S$} \ar^{0.05}[r] \ar^{0.95}[rd] & \fbox{$S$}\ar^{1-V}[r] \ar^{V}[rd]  & \fbox{$S_u$} \ar[r]  & \fbox{$E$}\ar^{0.5}_2[r] \ar^{0.5}_2[ru]  &  \fbox{$I_s$} \ar_5[r] \ar_2[d] & \fbox{$R$}   \\  
 & \fbox{$R$} & \fbox{$S_v$} \ar^{1-r_i}[ru]  &  & \fbox{$Q$} \ar[ur]_{14} } 
 \label{eq:SEIR}
\end{equation}
  \caption{Agent states with transition probabilities and average durations in days apearing as super- and subscripts. }\label{fig:SEIR}
\end{figure}
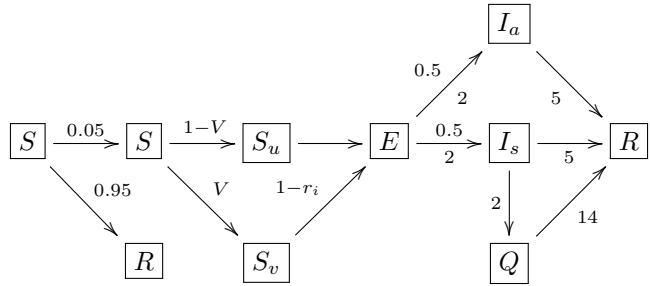

\begin{table}[h]
    \centering
    \begin{tabular}{ll|ll}
    \textbf{Parameter} & \textbf{Value} & \textbf{Parameter} & \textbf{Value} \\
    \hline 
    Incubation period     &  Geometric(3) days & Exogenous &  2 weekly \\
    Infectious period    &   Geometric(5) days & Antibody attribute &  0.20 \\
    Observation period &     Geometric(2) days &  Antibody protection &  0.95 \\
    Quarantine period &     14 days & Inward protection & $\{0.2, 0.5, 0.8\}$  \\
    $R_0$ &     3  & Outward protection & $\{0.2, 0.5, 0.8\}$
    \end{tabular}
    \caption{Infection parameters.}
    \label{tbl:par}
\end{table}

COVID-19 infections spread during meetings. Meetings include classes; the broad campus environment such as: hallways, elevators, lobbies, dining halls; time in dorms; clubs; and socializing. Each meeting occurs on a time and date and has a set \emph{duration} in minutes. Duration is a measure of risk-intensity rather than time elapsed during a meeting. We scale it by the number of agents in the meeting and the risk of infection spread in the meeting type. For example, socializing has a longer duration than class time, not because more time is spent doing so, but because it is a riskier setting for infection spread \cite{college_superspreader, cornell3}. Complete details regarding meeting structure are in Section~\ref{sec:meetings}.

Each individual in a meeting in the susceptible state acquires exposure time equal to the meeting duration times the number of attendees in the infected state also at the meeting. At the end of each day, the total number of exposure minutes for each susceptible agent is tallied. This total is scaled by the \emph{infection rate} which results in the probability the agent becomes infected on that day. The other manner in which infections occur in our model is through exogenous exposure in the non-campus community. We set the \emph{average exogenous exposures per week} by applying a fixed (small) probability of becoming infected to each agent at the end of each day.  Our model has 2 exogenous infections per week on average. Given our model's population of 16800 agents, this corresponds to 1.7 positive cases per 100,000 agents per day. At the time of writing, New York City has a rate roughly 10 times this, but we expect the rate to drop by Fall 2021 \cite{nyc_tracker}.

The infection rate in our model is set to obtain an \emph{average reproduction number} $R_0=3$, which represents the average number of infections caused by an exposed agent in an entirely susceptible population.
  Estimates for COVID-19 spread in large communities (such as cities and countries) put $R_0$ in the range $[2.0,3.0]$ \cite{rahman2020basic, abbott2020transmissibility, ueki2020effectiveness, zhou2020preliminary, zhao2020preliminary}, but there are some higher estimates \cite{sanche2020early}. The statistic varies by community contact structure. It has been observed that $R_0$ is larger in reopened universities \cite{college_superspreader}. For example, \cite{gressman2020} sets $R_0=3.8$ in their campus COVID-19 model for a mostly residential urban university.  See  \cite{gressman2020} and \cite{bahl2020modeling} for more discussion about elevated $R_0$ levels in a university setting.
  
\begin{figure}[h]
           \begin{floatrow}
             \ffigbox{\includegraphics[width = .49 \textwidth]{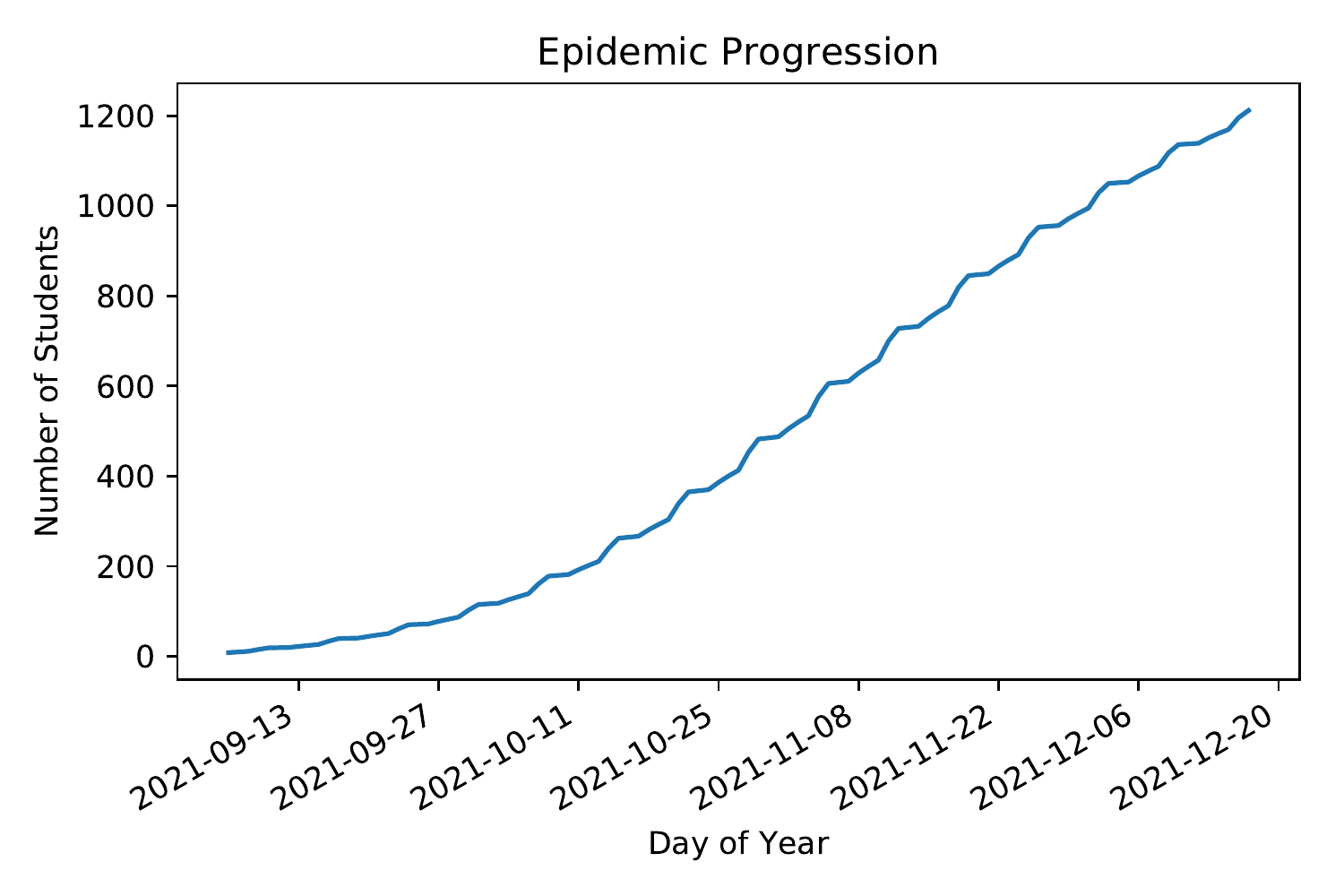}}{\caption{The number of infections over time in our base model with $R_0=3$ and no vaccination. }\label{fig:bm_curve}}
             \ffigbox{\includegraphics[width = .49 \textwidth]{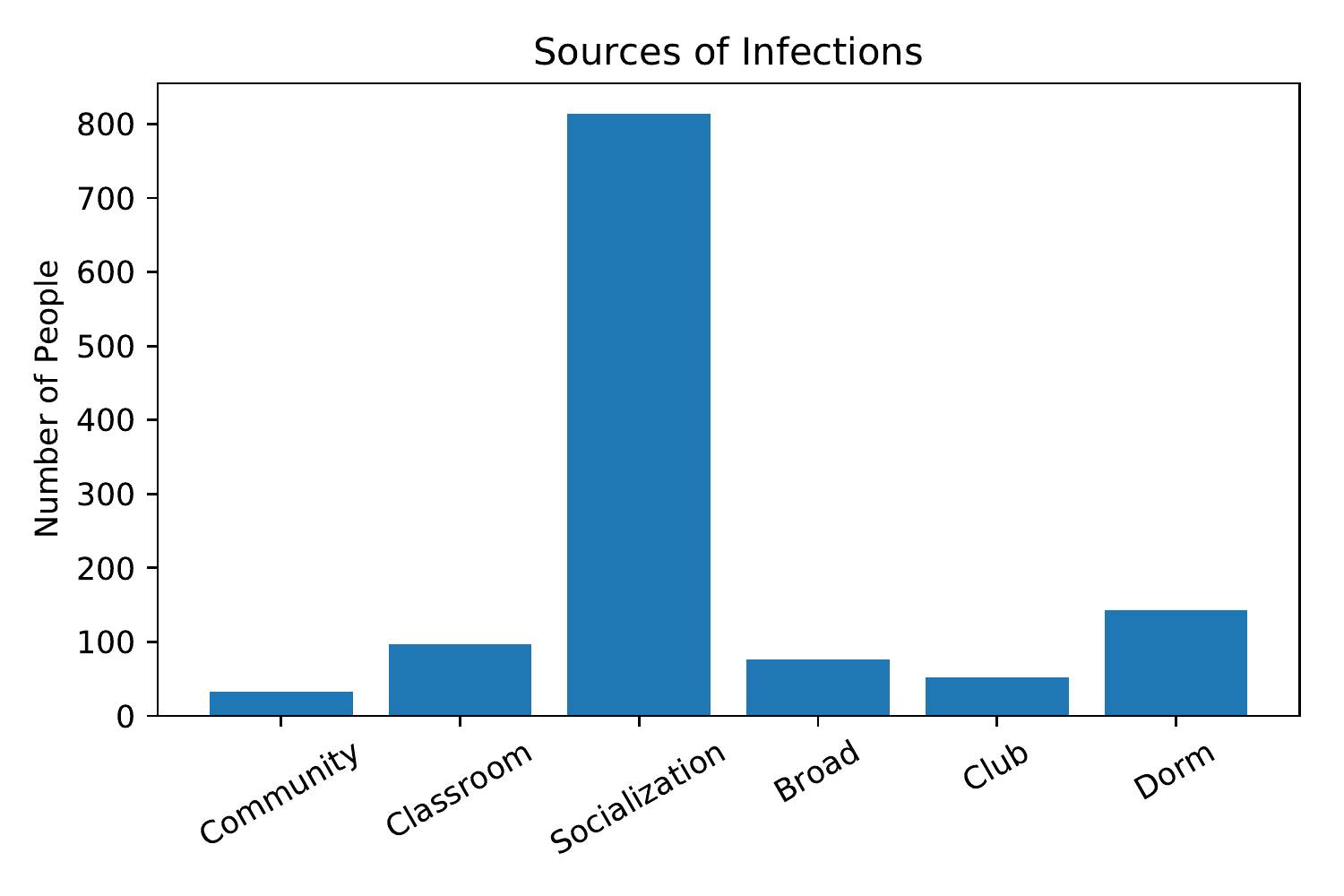}}{\caption{Total infections by source in our base model with $R_0=3$ and no vaccination.}\label{fig:bm_source}}
           \end{floatrow}
\end{figure}

  The main statistic we consider is the total number of agents ever in the exposed state over the course of a 15-week semester. We refer to this as \emph{total infections}. 
  The model is initiated with 10 randomly selected students in the exposed state. Our \emph{base model} represents a reopening with the full population present on campus, antibodies present in 20\% of the population, but no vaccination and no screening testing. We assume that facemasks are used except when socializing in private. This is accounted for by lowering the risk of infection spread in public spaces such as classrooms and broad environment. In the base model there is no active monitoring of the number of cases, so no adaptive policies (such as temporary suspension of in-person instruction) are ever implemented.
  With $R_0=3$, we find, on average, 1200 total infections in the base model with no vaccination and 20\% antibody incidence. Figures~\ref{fig:bm_curve}  gives a sense of how the number of infections evolves over time. Note that we do not include a ``Thanksgiving Effect" with a November rise in infections in our model. Figure~\ref{fig:bm_source} shows the average number of infections occurring in each setting. As mentioned previously, socializing is the main venue for infection spread in our model. Note that we have 400 students living in the residential dorms. This is in alignment with Baruch College and amounts to less that 2\% of the student population.

Vaccination and antibody status impact agents' susceptibility and infectiousness. Each agent is assigned the \emph{vaccinated attribute} independently with probability $V$. Such agents have \emph{inward protection} factor $r_i$ and \emph{outward protection} factor $r_o$. Vaccinated agents contribute a factor of $1-r_o$ of exposure time to susceptible agents in each meeting they are present at. When computing the probability vaccinated agents are infected at the end of a day, the probability is multiplied by $1-r_i$. All COVID-19 infections in vaccinated agents are classified as asymptomatic.  Our medium-effectiveness scenarios set $r_i = 0.5 = r_o$, while low-effectiveness has $r_i=0.2=r_o$ and high-effectiveness has $r_i = 0.8 = r_o$. We perform a sensitivity analysis to setting $r_i \neq r_o$ in Section \ref{sec:sensitivity}. Table \ref{tbl:par} shows the relevant infection parameters. Once exposed, vaccinated agents progress through the stages of COVID-19 infection (exposed, infectious, recovered) as a normal susceptible agent would. See Figure~\ref{fig:SEIR} for a schematic.



\subsection{Meeting Structure} \label{sec:meetings}

Each of the 16000 student is assigned a \emph{Year} in 1,2,3,4 (in equal proportions) and an \emph{Area} in Business,  STEM, and Humanities. The proportion in each area is $75\%,15\%,$ and $10\%$, respectively. The $800$ faculty are divided in the same proportions as students to each of the three areas. 
The student/faculty designation, year and area of an agent play a role in the meetings they attends. Broadly, there are five types of meetings: class, broad, club, social, and residential. Students and faculty interact in class time and broad meetings. Only students interact in club, social, and residential meetings.

\subsubsection{Courses}
Courses meet twice per week either MW or TuTh for $c \cdot 100/L$ minutes each class where $c =1/10$ is a scaling parameter to account for reduced transmission probability in classrooms and $L$ is the number of students enrolled. Courses are either General Interest (G), Business (B), STEM (S), or Humanities (H). Each class is independently designated as either a MW or TTh meeting class with probability $1/2$ each.  The number of classes of various sizes in Table~\ref{tbl:classes} are chosen so that 20\% of all classes are General Interest, and the proportions of classes of each size 
align with the counts provided by the Baruch College Common Data Set.

\begin{table}[h]
\centering 
\begin{tabular}{l|rrrrrrr|l}
\textbf{Class Size}   & \textbf{10} & \textbf{20} & \textbf{30} & \textbf{40} & \textbf{50} & \textbf{75} & \textbf{150} & \multicolumn{1}{l}{$(C_X,T_X)$} \\
\hline 
G               & 8           & 40          & 140         & 60          & 30          & 40          & 10           & (328, 13480)                     \\
B               & 24          & 120         & 420         & 180         & 90          & 120         & 30           & (984, 40440)                \\
S               & 5           & 24          & 84          & 36          & 18          & 24          & 6            & (196, 8088)                      \\
H               & 3           & 16          & 56          & 24          & 12          & 16          & 4            & (131, 5392)              \\
\hline
\textbf{Total} & 40          & 200         & 700         & 300         & 150         & 200         & 50           & (1640, 67400)              
\end{tabular}
\caption{Counts for various class sizes.}\label{tbl:classes}
\end{table}



We draw inspiration for how class meetings are generated in \cite{gressman2020} using enrollment histogram data and order statistics to create correlations among courses in students among different years.  Let $C_{X,y}$ be the total number of classes of size $y$ in area $X \in \{B,S,H,G\}$. For example, $C_{B,30} = 420$. Let $C_X = \sum_{y} C_{X,y}$.  Let $\vec X = ( X_1, \hdots , X_{C_{X}})$ be the sizes of classes in area $X$ arranged from largest to smallest. For example $$\vec S = (\underbrace{150, 150, \hdots, 150}_{C_{S,150}},\underbrace{ 75, 75,  \hdots , 75}_{C_{S,75}}, 50, \hdots, 10).$$ 
Let $T_X = \sum_{i=1}^{C_X} X_i$ be the total number of seats offered across all of the courses in area $X$.
Form the vector 
$$\vec p_X = (p_1(X), \hdots, p_{C_X}(X)) \text{  with  }
p_i(X) = \f{X_i}{T_X}.$$

Index the courses in $\vec G \oplus \vec X := (G_1,\hdots, G_{C_G}, X_1,\hdots, X_{C_X})$ as $\Omega_{X} = \{1,2,\hdots, C_{G}+C_X\}$. Define the random variable $Y(X)$ that takes values in $\Omega_X$ where, with probability $1/5$, $Y(X)$ is drawn from a multinomial with distribution $\vec p_G$ on $1,\hdots, C_G$ and, with probability $4/5$, is drawn from a multinomial with distribution $\vec p_X$ on $C_G+1,\hdots, C_X$. 
We then assign classes to four students in area $X$, one of each year, simultaneously by sampling four independent $Y_1(X),\hdots, Y_4(X) \sim Y(X)$. Let $$Y_{(1)}(X) \leq Y_{(2)}(X) \leq Y_{(3)}(X) \leq Y_{(4)}(X)$$ be the arrangement of the $Y_k(X)$ from least to greatest.  The student in year $k$ is assigned class $Y_{(k)}(X)$. Each student is assigned four classes in this manner.  

This construction ensures that the amount of each class size in each area is proportional to the ratios in Table~\ref{tbl:classes}. Using order statistics ensures that students in an earlier year are more likely to take large, general interest classes. 
Faculty in the corresponding area are assigned to teach two uniformly samples courses in their area.

\subsubsection{Broad environment}
{All agents spend $20/L$ minutes per M, T, W, Th meeting with the $L$ students and faculty in their area, and $10/16800$ total minutes per week meeting with all agents in the model. This represents ambient environmental contacts (hallways, elevators, lobbies, gym, library) that occur on campus. 
}

\subsubsection{Clubs}
{Clubs} meet $100/L$ minutes on Thursday where $L$ is the size of the club. There are 50 General Interest, 30 Business, 20 STEM, and 10 Humanities clubs. Each student joins a uniformly random general interest club with probability $1/5$ and a uniformly random club in their area with probability $1/5$. The probability a student does not participate in any clubs is $(4/5)*(4/5) = 16/25 = 0.64$. This is in line with the participation rates for clubs at Baruch College according to the 2018 Student Experience Survey.

\subsubsection{Residence Hall}
{
Pick 400 total students uniformly at random from years 1 and 2 to live in residence halls. Pair these students up into 200 groups of two students each representing roommates. Each roommate group meets 300 minutes per day. The entire group of 400 students in the residence hall spend 100/400 minutes together per week. 
}

\subsubsection{Social}

Small and large social groups are formed via a Markov process. All students are labeled as low, medium, or high socializers. In line with socializing surveys from \cite{freshman2019}, the probability a student is a low socializer is 0.15, medium is 0.45, and high is 0.40.

Let $\L, \M,$ and $\H$ be the sets of low, medium, and high socializers. Furthermore, let $X_k(Y)$ be the set of level $Y$ socializers from area $X$ in year $k$. For example, $H_3(\M)$ are medium-socializers in their third year of humanities.
Whenever a student is sampled from a group $Z$, the sampling is done so that the student is uniformly sampled from $Z \cap \L$ with probability $0.10$, from $Z \cap \M$ with probability $0.30$, and from $Z \cap \H$ with probability $0.60$. Call this method $(*)$.

A \emph{small social group} is formed according to the following algorithm.

\begin{enumerate}[label = (\roman*)]
    \item Select a student from the entire population according to $(*)$. Suppose they are from area $X$ and year $k$.
    \item The next student is sampled according to $(*)$ from:
        \begin{itemize} 
            \item The entire student population with probability $1/6$. 
            \item All students in year $k$ with probability $1/6$.
            \item All students in area $X$ with probability $1/6$.
            \item All students in area $X$ and year $k$ with probability $1/2$. 
        \end{itemize}
    \item With probability $1/2$, no more members are added to the group.  With probability $1/2$ the algorithm continues using the year and area of the newly added member to generate the next choice via (ii) and (iii). 
\end{enumerate}
A \emph{medium social group} is formed by replacing the probability of adding an additional member to the group at step (iii) with $9/10$. Every Friday there is a \emph{large social group} consisting of five uniformly randomly selected medium social groups. The duration is $2000 / L$ minutes $L$ the total number of people in the meeting. The long duration of large social groups is capturing the ``superspreader phenomenon" observed on campuses during the 2020-2021 school year \cite{college_superspreader}.

Small social groups have expected size 3. These model close friends who study, eat, and pass time together. Medium social groups have expected size 11 and large social groups have expected size 55. These model larger social gatherings such as parties or events. 

{Each small group meets with probability 1/2 on each weekday M, Tu, W, Th for $1000/L$ minutes where $L$ is the size of the group. This makes a minute of socializing ten times higher risk than a usual minute. Each medium group meets with probability $1/2$ on Th and F for $1000/ L$ minutes where $L$ is the number of people in the meeting. Large social groups meet for $1000/L$ minutes on F (with probability 1). These random choices are made for the first week and repeated for all weeks thereafter. The parameter $s$ scales for the higher risk of infection transmission during socializing since facemasks and social distancing are less likely to be employed. 100 is chosen so that the scaling is relative to the meeting time of a course.}
We form 3000 small social groups, 300 medium social groups, 50 large social groups for the base model. 

\subsection{Additional Sensitivity Analysis} \label{sec:sensitivity}

\begin{figure}
           \begin{floatrow}
             \ffigbox{ \includegraphics[width = .49 \textwidth]{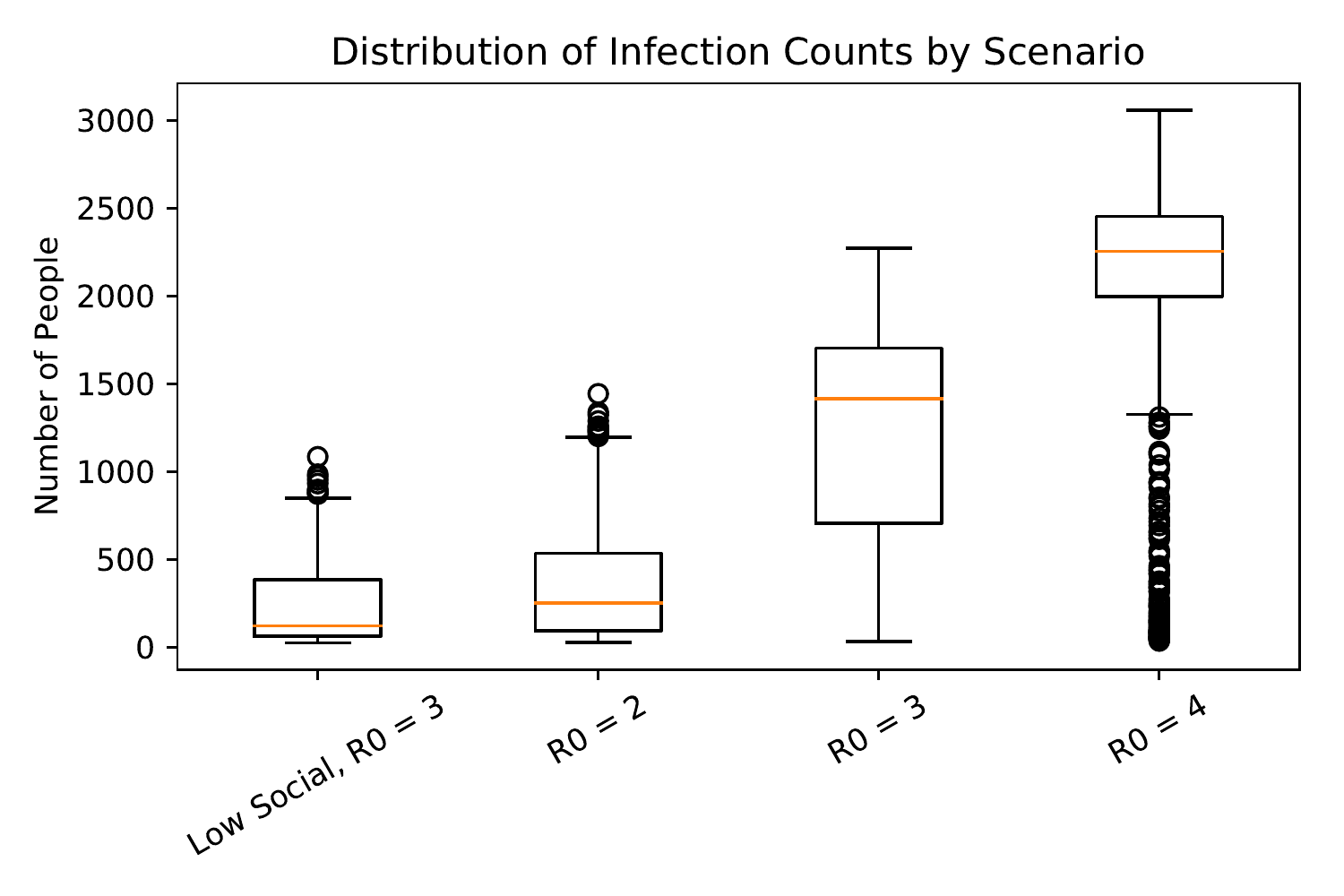}}{\caption{Total infections with $R_0=3$ and socializing as in the base model as well as the base model with $R_0=2,3,4$. The data is obtained from 1000 runs of each scenario. }\label{fig:bms1}}
             \ffigbox{\includegraphics[width = .49 \textwidth]{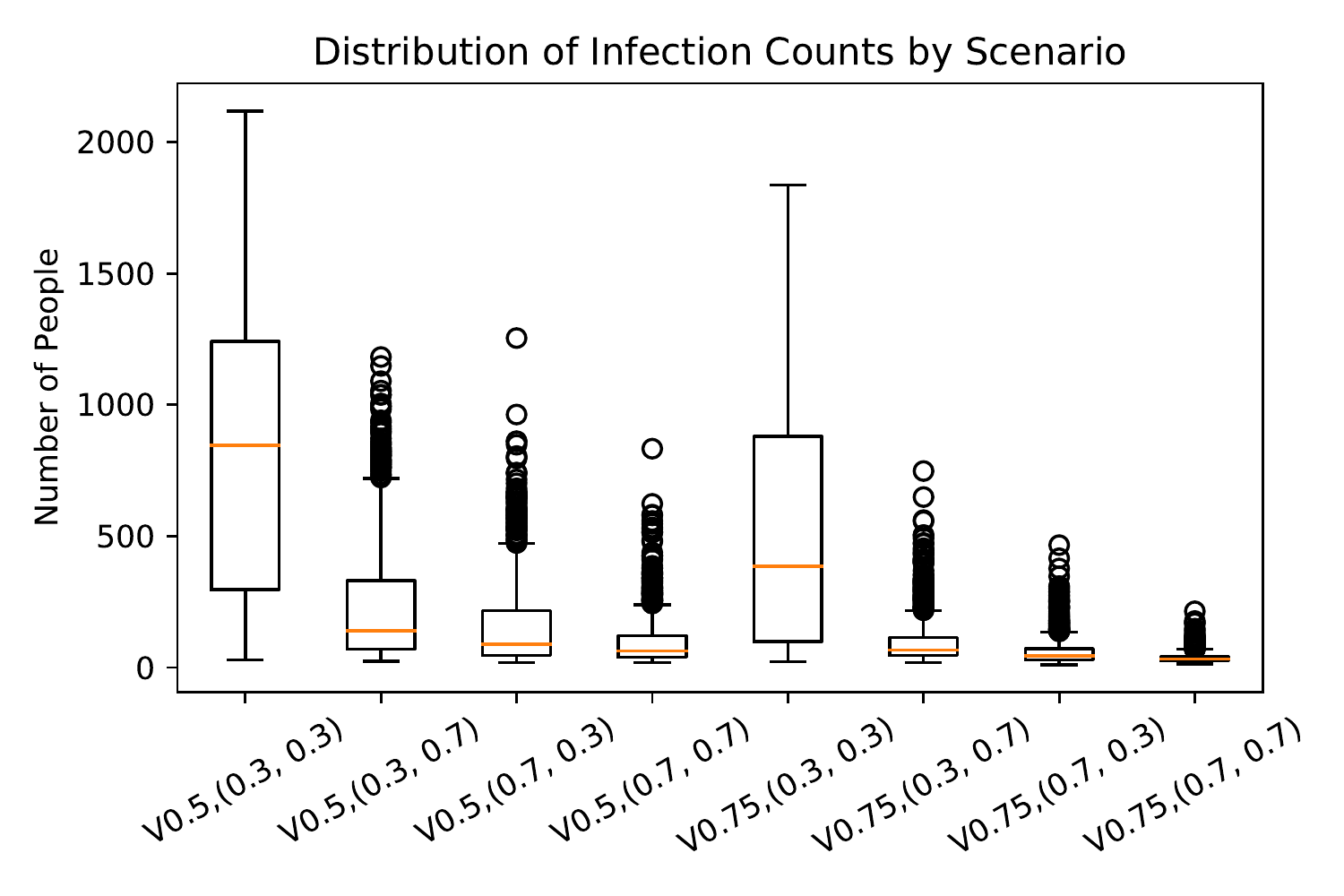}}{\caption{ Total infections in the base model with different combinations of $r_i$ and $r_o$.}\label{fig:bms2}}
           \end{floatrow}
\end{figure}

The reproduction number is a phenomenological output of the infection biology and contact structure in the model. Thus, it is is difficult to  calibrate in heterogeneous populations (see the discussion in \cite{bahl2020modeling}). For this reason, we additionally run our base model with $R_0=2$ and $R_0=4$. Since socialization is a major source of infection spread, we also include a version with $R_0=3$ and half as much social interactions. These variations are displayed in Figure~\ref{fig:bms1}. As expected, total infections are greatly reduced by decreasing socializing. Moreover, we see that total infections are sensitive to our choice of $R_0$. This is more reason for administrators to exercise caution in their reopening plans. Lastly, Figure~\ref{fig:bms2} shows box plots for total infections with unequal vaccine effectiveness parameters $(r_i,r_o) \in \{ 0.3, 0.7\}^2$. We find that the impact from each is roughly the same. This suggests that our choices of setting $r_i=0.5=r_o$ in our main analysis and also $r_i = r_o$ in our low-, medium- and high-effectiveness vaccine scenarios reasonable simplifications to make.

   \subsection{Code Access}
The code for the project is publicly available on Github at the address
\begin{center}
\href{https://github.com/MAS-Research/SEIR-Campus}{https://github.com/MAS-Research/SEIR-Campus}
\end{center} 
as an extension of \cite{zalesak2020seir}.  There are two files associated with this paper:  \emph{CunyCovid.ipynb} replicates the simulations discussed in this paper and writes the data to a file.  \emph{ImageRendering.ipynb} loads the simulation data to produce the graphics shown in this paper.

\end{document}